\documentclass[12pt,letterpaper]{article}

\usepackage{amsmath,amsthm}
\usepackage{fullpage}
\usepackage[dvips]{epsfig}
\usepackage{epsf}
\usepackage{float}

\newtheorem{theorem}{Theorem}
\newtheorem{corollary}[theorem]{Corollary}
\newtheorem{lemma}[theorem]{Lemma}

\title{On NFAs Where All States are Final, Initial, or Both}

\author{Jui-Yi Kao,\footnote{Author's current address:  Department of
Computer Science, Stanford University, Stanford, CA  94305, USA,
{\tt erickao@stanford.edu}.}
\ Narad Rampersad\footnote{Author's current address:  
Department of Mathematics and Statistics,
University of Winnipeg,
515 Portage Avenue,
Winnipeg, MB R3B 2E9,
Canada,
{\tt narad.rampersad@gmail.com}.}, 
\ and Jeffrey Shallit \\
School of Computer Science\\
University of Waterloo\\
Waterloo, Ontario  N2L 3G1\\
Canada\\
{\tt shallit@cs.uwaterloo.ca}}

\begin{document}
\date{\today}
\maketitle

\def\union{\ \cup \ }
\def\intersect{\ \cap \ }
\def\pref{{\rm{pref}}}
\def\suff{{\rm{suff}}}
\def\subword{{\rm{fact}}}

\begin{abstract}
We examine questions involving
nondeterministic finite automata where all states are final, initial, or
both initial and final.
First, we prove hardness results for the nonuniversality
and inequivalence problems for these NFAs.   Next, we characterize the languages
accepted.  Finally, we discuss some state complexity problems involving such
automata.
\end{abstract}

\section{Introduction}
\label{intro}

Nondeterministic finite automata (NFAs) differ from deterministic finite
automata (DFA) in at least two important ways.  First, they can be 
exponentially more concise in expressing certain languages, as it is
known that there exist NFAs on $n$ states for which the smallest equivalent
DFA has $2^n$ states \cite{Ershov:1962,Moore:1971,Meyer&Fischer:1971}.  Second, while it is possible to test inequivalence
and nonuniversality for DFAs in polynomial-time, the corresponding problems
for NFAs are PSPACE-complete \cite[Lemma 2.3, p.\ 127]{Meyer&Stockmeyer:1972}.

In this paper, we consider NFAs with certain natural restrictions, such 
as having all states final, all states initial, or all states both
initial and final.  Although imposing these conditions 
significantly narrows the class of languages accepted (see \S~\ref{charac}),
we show that there is still an exponential blow-up in converting to an
equivalent DFA, and the corresponding decision problems are still
PSPACE-complete.  Furthermore, these restricted NFAs are intimately related to
languages that are prefix-closed, suffix-closed, or factor-closed
(see \S~\ref{charac} and \cite{Brzozowski:2008}), and have close
connections with certain decision questions on infinite words and
a decision problem on Boolean matrices
\cite{Rampersad:2009}.

     Here is a brief outline of the paper.
In Section~\ref{notation}, we give some basic definitions and
notation.  In Sections~\ref{hardness}, \ref{hardnessi}, and
\ref{hardnessif}, we prove an assortment of hardness results on NFAs
with various restrictions on their initial and final states.    In
Section~\ref{charac}, we give a simple characterization of the
languages accepted by NFAs with these restrictions.  In
Sections~\ref{statecomp} and \ref{statecomp2}, we give our main
results, on state complexity.  We end with Sections~\ref{complement}
and \ref{shortest}, where we discuss the complexity of complement and
the length of the shortest word not accepted.

\section{Definitions and notation}
\label{notation}

We recall some basic definitions.  For further details,
see \cite{Hopcroft&Ullman:1979}.
A \emph{non-deterministic finite automaton} (NFA) $M$ is
a quintuple $M = (Q,\Sigma,\delta,I,F)$, where $Q$ is a finite
set of states; $\Sigma$ is a finite alphabet;
$\delta\,:\,Q \times \Sigma \to 2^Q$ is the transition function,
which we extend to $Q \times \Sigma^*$ in the natural way; $I \subseteq Q$
is the set of initial states\footnote{In the ``formal'' definition of an NFA 
(e.g., \cite[p.\ 20]{Hopcroft&Ullman:1979}), only one
initial state is typically allowed.  However, NFAs with multiple initial
states can clearly be simulated by NFA-$\epsilon$'s, and hence by NFAs with
at most one more state.  We find
it useful to avoid the complication of $\epsilon$-transitions and
simply allow having multiple initial states here.};
and $F \subseteq Q$ is the set of final states.
An NFA $M$ accepts a word $w \in \Sigma^*$ if
$\delta(I,w) \cap F \neq \emptyset$.
The language of all words accepted by $M$ is denoted $L(M)$.

A \emph{deterministic finite automaton} (DFA) $M$ is defined as an NFA
above, with the following restrictions: $M$ has only one initial state
$q_0$, and $|\delta(q,a)| = 1$ for all $q \in Q$ and $a \in \Sigma$.

The \emph{state complexity} of a regular language $L$ is the number of
states in the minimal DFA accepting $L$.
Given an operation on regular languages, we also define
the state complexity of that operation to be the number of states that
are both sufficient and necessary in the worst-case for a DFA to
accept the resulting language.

\section{Hardness results}
\label{hardness}

       First, we discuss the case of NFAs with a single initial state,
and where all states are final.

       Consider the following decision problem

\bigskip

\noindent{\tt NFA-INEQUIVALENCE-ASF}($k$):  Given two NFAs $M_1$ and $M_2$,
over an alphabet with $k$ letters,
each having the property that all states are final states,
is $L(M_1) \not= L(M_2)$?

\bigskip

       We will prove the following theorem.

\begin{theorem}
       {\tt NFA-INEQUIVALENCE-ASF} ($k$) is PSPACE-complete
for $k \geq 2$, but solvable in polynomial time for $k = 1$.
\label{first}
\end{theorem}

\begin{proof}
       First, let us consider the case where $k = 1$.  Let $M$
be a unary NFA (over the alphabet $\Sigma = \lbrace a \rbrace$)
with all states final.  Then $L(M)$ is either finite
or $\Sigma^*$, depending on whether there is a cycle in the
directed graph $G$ given by the transitions of $M$.  Furthermore,
if $M$ has $n$ states, then $a^n \in L(M)$ iff $G$ has a cycle reachable
from $q_0$.
Therefore, we can determine $L(M)$ efficiently by checking first
if $a^n$ is accepted.  If it isn't, we then successively check
whether $a^{n-1}$, $a^{n-2}, \ldots, a^1, \epsilon$ are accepted.
If the first string in this list that is accepted is $a^i$, then
$L(M) = \lbrace \epsilon, a, \ldots, a^i \rbrace$.  Thus we can
check whether $L(M_1) \not= L(M_2)$ efficiently.


The {\tt NFA-INEQUIVALENCE-ASF} problem is in PSPACE, since the more general
{\tt NFA-INEQUIVALENCE} problem (problem AL1 in Garey and Johnson
\cite[p.\ 265]{Garey&Johnson:1979}) is well-known to be in PSPACE.  A proof
of this result can be found in Sipser \cite[p.\ 315]{Sipser:1996}.

       Now we need to see that {\tt NFA-INEQUIVALENCE-ASF} is
PSPACE-hard.  To do so, we consider the specialization
{\tt NFA-NONUNIVERSALITY-ASF}:

\bigskip

\noindent{\tt NFA-NONUNIVERSALITY-ASF}($k$):  Given an 
NFA $M$ over an alphabet $\Sigma$ with $k$ letters,
having the property that all states are final states,
is $L(M) \neq \Sigma^*$?

\bigskip

Clearly, if we prove the stronger result that
{\tt NFA-NONUNIVERSALITY-ASF}($k$) is PSPACE-hard,
then it will follow that {\tt NFA-INEQUIVALENCE-ASF}($k$) is PSPACE-hard,
by choosing one of the NFAs to be the one-state NFA with a loop back
to the single state on every input symbol.  So it will suffice to prove
the following lemma:

\begin{lemma}
{\tt NFA-NONUNIVERSALITY-ASF}($k$) is PSPACE-hard for $k \geq 2$.
\end{lemma}

\begin{proof}

First, let us consider the case where $k \geq 3$.  We
reduce from the following decision problem, which is well-known
to be PSPACE-complete \cite[p.\ 265]{Garey&Johnson:1979} for
$k \geq 2$:

\bigskip

\noindent{\tt NFA-NONUNIVERSALITY}($k$):  Given an NFA
$M$ over an alphabet $\Sigma$ with $k$ letters, is $L(M) \not= \Sigma^*$?

\bigskip

      Here are the details of the reduction.

      Given an NFA $M$ over an alphabet of size $k$,
we transform it to an NFA $M'$ with all states
final, over an alphabet of size $k+1$, as follows:
$M'$ is identical to $M$, except that we add a transition from
each final state of $M$ to the initial state $q_0$ on a new
symbol, say $\#$, and then we change all states to be final states.
This construction is illustrated below in Figure~\ref{nfa4}.

\begin{figure}[htb]
\centering
\includegraphics[scale=0.75]{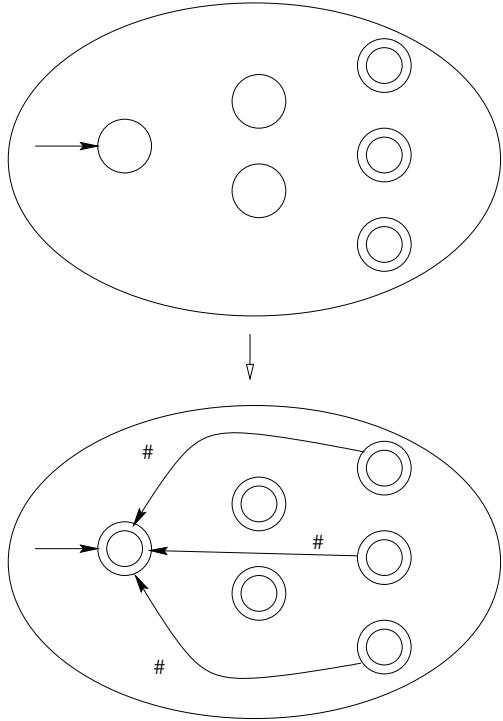}
\caption{The transformation of $M$ to $M'$.}
\label{nfa4}
\end{figure}

      Let $\Delta = \Sigma \ \cup \ \lbrace \# \rbrace$.

      We now claim that $L(M) \not= \Sigma^*$ iff 
$L(M') \not= \Delta^*$, or, equivalently,
$L(M) = \Sigma^*$ iff $L(M') = \Delta^*$.

     Suppose $L(M) = \Sigma^*$.  Then for each string $w \in \Sigma^*$,
there exists a final state $p(w)$ of $M$ such that
$p \in \delta(q_0, w)$.  Let $x \in \Delta^*$.  If $x \in \Sigma^*$,
the result is clear.  Otherwise write
$x = x_1 \# x_2 \# x_3 \# \cdots \# x_n$, where each $x_i \in \Sigma^*$.
Now there exists an accepting computation for $x$ in $M'$, which
starts in $q_0$, follows $x_1$ to the state $p(x_1)$, then follows
the transition on $\#$ back to $q_0$ of $M'$, then follows
$x_2$ to $p(x_2)$, etc.   Thus $L(M') = \Delta^*$.

      Now suppose $L(M') = \Delta^*$.  Then, in particular, $M'$ accepts
all strings of the form $w \#$ where $w \in \Sigma^*$.  In order
for $M'$ to accept $w \#$, it must be the case that there is a transition
from a state $p \in \delta(q_0,w)$ on $\#$ in $M'$.  But then this state
is final in $M$, by construction, so $w$ is accepted by $M$.  Thus
$L(M) = \Sigma^*$.

      This completes the reduction.  Note that our construction increases
the size of the alphabet by $1$, so that we have shown that\\
\centerline{ {\tt NFA-NONUNIVERSALITY}($k$) reduces to {\tt NFA-NONUNIVERSALITY-ASF}($k+1$).} \\
Since {\tt NFA-NONUNIVERSALITY}
is PSPACE-hard for $k \geq 2$, we have proved the lemma for $k \geq 3$.

      It remains to show {\tt NFA-NONUNIVERSALITY-ASF}($2$) is PSPACE-hard.
To do this, we show by recoding that 
{\tt NFA-NONUNIVERSALITY-ASF}($4$) reduces to
{\tt NFA-NONUNIVERSALITY-ASF}($2$).

Here are the details.  Given a machine $M$ over the input alphabet
$\Sigma = \lbrace 0, 1, 2, 3 \rbrace$ with all states final, we create
a new machine $M'$ over the input alphabet $\Delta = \lbrace 0, 1 \rbrace$.
Each transition out of a state $A$ is recoded, and two new final
states are introduced, so that
\begin{itemize}

\item a transition on $0$ is replaced by a transition on $0$ followed by $0$
\item a transition on $1$ is replaced by a transition on $0$ followed by $1$
\item a transition on $2$ is replaced by a transition on $1$ followed by $0$
\item a transition on $3$ is replaced by a transition on $1$ followed by $1$
\end{itemize}
See Figure~\ref{nfa5}.

\begin{figure}[H]
\centering
\includegraphics[scale=0.75]{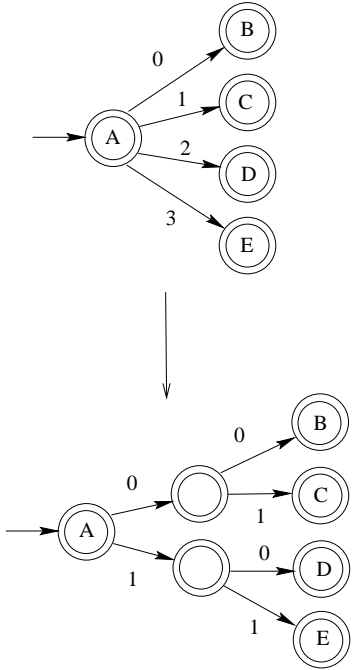}
\caption{The transformation of $M$ to $M'$.}
\label{nfa5}
\end{figure}

We claim that $L(M) = \Sigma^*$ iff $L(M') = \Delta^*$. 
\end{proof}

This completes the proof of Theorem~\ref{first}.
\end{proof}

\begin{corollary}
Minimizing an NFA with all states final, over an alphabet of 
size $\geq 2$, is PSPACE-hard.
\end{corollary}

\begin{proof}
If we could minimize an NFA with all states final, we could also solve
the nonuniversality problem $L(M) \not= \Sigma^*$ as follows:  first
we minimize the NFA.  If it has $\geq 2$ states, we say ``yes''.
Otherwise we inspect the transitions (if any) of the minimized NFA,
and check if the single state is final and that there is a loop on
every element of the alphabet.  If so, we say ``no''; otherwise, we
say ``yes''.
\end{proof}

\section{Generalized NFA with all states initial}
\label{hardnessi}

Now we consider a variant of the problems considered in 
Section~\ref{hardness}.  These variants
concern generalized NFAs with multiple
initial states allowed, in which all states are initial states and there
is only one final state.
We consider the following decision problems:

\bigskip

\noindent{\tt NFA-INEQUIVALENCE-ASI}($k$):  Given two NFAs $M_1$ and $M_2$,
over an alphabet with $k$ letters,
each having the property that all states are initial states and only
one state is final, is $L(M_1) \not= L(M_2)$?

\bigskip

\noindent{\tt NFA-NONUNIVERSALITY-ASI}($k$):  Given an 
NFA $M$ over an alphabet $\Sigma$ with $k$ letters,
having the property that all states are initial states and only one
state is final,
is $L(M) \neq \Sigma^*$?

\bigskip

We prove the following theorem.

\begin{theorem}
Both {\tt NFA-INEQUIVALENCE-ASI}($k$) and {\tt NFA-NONUNIVERSALITY-ASI}($k$)
are PSPACE-complete for alphabet size
$k \geq 2$, but solvable in polynomial time for
$k = 1$.   Furthermore, minimizing an NFA with all states initial and
one state final is PSPACE-hard for $k \geq 2$.
\end{theorem}

\begin{proof}
These results follow trivially from the results in the previous section
by observing that $L$ is accepted by an NFA $M$ with a single initial state
and all states final iff $L^R$ (the language formed by reversing all the
strings of $L$) is accepted by $M^R$, the generalized NFA formed by
reversing all the transitions of $M$, and changing initial states into
final and vice versa. 
\end{proof}

\section{All states both initial and final}
\label{hardnessif}

Our original motivation in Section~\ref{intro} involved
generalized NFAs where all states are both
initial and final.
Consider the following decision problem:

\bigskip

\noindent{\tt NFA-INEQUIVALENCE-ASIF}($k$):  Given two NFAs $M_1$ and $M_2$,
over an alphabet with $k$ letters,
each having the property that all states are both initial and final,
is $L(M_1) \not= L(M_2)$?

\bigskip

\begin{theorem}
{\tt NFA-INEQUIVALENCE-ASIF} ($k$) is PSPACE-complete
for $k \geq 2$, but solvable in polynomial time for $k = 1$.
\label{asif}
\end{theorem}

\begin{proof}
The idea is similar to that in the proof of Theorem~\ref{first}.
We only indicate
what needs to be changed.

Once again, we work with the ``easier'' problem

\bigskip

\noindent{\tt NFA-NONUNIVERSALITY-ASIF}($k$):  Given an NFA $M$
over an alphabet with $k$ letters,
having the property that all states are both initial and final,
is $L(M) \neq \Sigma^*$ ?

\bigskip

We can show that {\tt NFA-NONUNIVERSALITY}($k$) reduces to
{\tt NFA-NONUNIVERSALITY-ASIF}($k+1$) using a simple variant of our previous
proof. 
Given $M$, an NFA over an alphabet $\Sigma$ of $k$ symbols, we modify it to
obtain $M'$, an NFA over an alphabet $\Delta = \Sigma \ \cup \ \lbrace
\# \rbrace$ of $k+1$ symbols, as follows.
First, we delete all states of $M$ not reachable from $q_0$, the start
state.  Next, we introduce a new
symbol $\#$ and transitions on $\#$ from each of the final states of $M$ to 
$q_0$.  Finally, we change all states to be both initial
and final. 
We claim that $L(M) = \Sigma^*$ iff $L(M') = \Delta^*$.

The direction $L(M) = \Sigma^* 
\implies L(M') = \Delta^*$ is exactly as
before.  For the other direction, suppose $L(M') = \Delta^*$.  Then,
in particular,
for all $x \in \Sigma^*$, the string $\# x \#$ is accepted by $M'$.
Consider an accepting path for this string in $M'$.
It starts at some state (since all states are initial)
and then follows a transition on $\#$ to $q_0$.
The machine $M'$
now processes $x$ and arrives at some state $q$.  In order
for $M'$ to reach a final state on the last symbol,
$\#$, there must be a transition on $\#$
from $q$ to $q_0$.  But this can only be the case if $q$ was final in $M$.
Thus we have found an accepting path for $x$ in $M$, and so $L(M) = \Sigma^*$.

Thus we have shown {\tt NFA-NONUNIVERSALITY-ASIF}$(k)$ is PSPACE-complete
for $k \geq 3$, and thus, that {\tt NFA-INEQUIVALENCE-ASIF}$(k)$ 
is PSPACE-complete for $k \geq 3$.

To complete the proof of the theorem, we prove the following lemma.

\begin{lemma}
{\tt NFA-NONUNIVERSALITY-ASIF}$(2)$ is PSPACE-complete.
\end{lemma}

\begin{proof}
It is enough to show that {\tt NFA-NONUNIVERSALITY-ASF}$(3)$ reduces
to \break
{\tt NFA-NONUNIVERSALITY-ASIF}$(2)$.  The reduction has several steps,
but the basic idea is simply to recode the 3-letter alphabet
$\lbrace 0, 1, 2 \rbrace$
into strings over a
$2$-letter alphabet $\lbrace 1, 10, 100 \rbrace$.

Given an NFA $M$ with input alphabet $\Sigma = \lbrace 0,1,2 \rbrace$,
a single initial state $q_0$, and all states final, we first
modify $M$ to enforce the condition
that there be no transitions entering the initial state.  To do this, we double the
initial state, adding a new state $p_0$ with the same outgoing transitions as $q_0$,
and make any transitions formerly entering $q_0$ to enter $p_0$ instead.

Second, we enforce the condition that the labels of all transitions entering a particular
state be the same.  To do this, we triple each state except the initial state (which,
by construction, now has no incoming transitions), copying the outgoing transitions, and
assigning an incoming transition of each element of $\Sigma$ to one of the three states,
appropriately.

Third, we recode the transitions of the NFA, as follows:
\begin{eqnarray*}
0 \ \text{gets recoded as}\ 1 \\
1 \ \text{gets recoded as}\ 10  \\
2 \ \text{gets recoded as}\ 100 
\end{eqnarray*}
Of course, this recoding necessitates introducing intermediate states for transitions
on $1$ and $2$.  We call these intermediate states ``new'' and all other states
``old''.

The incoming transitions of each old state have the same labels, which
are either $1$, $10$, or $100$.  In our fourth step,
we add additional outgoing transitions, and states,
as depicted in Figure~\ref{transf}.   The dotted transitions indicate transitions
that include some nondepicted states, and the dashed circles indicate the additional
states added.
The effect of these additional transitions is to allow, from each old state
with an incoming arrow,
a path labeled by $1$ and then 3 or more zeroes that returns to $q_0$.


\begin{figure}[H]
\centering
\includegraphics[scale=0.8]{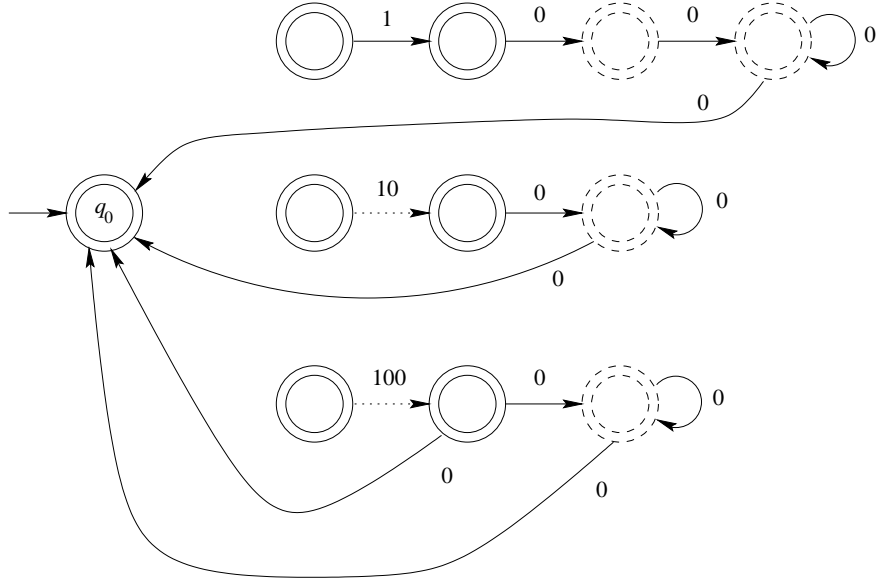}
\caption{Additional outgoing transitions}
\label{transf}
\end{figure}

Finally, we make all states both initial and final.
Call the resulting generalized
NFA $M'$.  We claim $M$ accepts $\Sigma^*$ iff $M'$ accepts $\Delta^*$, where
$\Delta = \lbrace 0,1  \rbrace$.  Define the morphism $h$ by
$0 \rightarrow 1$, $1 \rightarrow 10$, and $2 \rightarrow 100$.

Suppose $M$ accepts $\Sigma^*$.  We need to show that every $s \in \Delta^*$
is accepted by $M'$.  Let us identify the maximal blocks of $3$ or more
zeroes in $s$, if they exist.  These blocks either mark the beginning
or end of $s$, or else are bounded on the left by a string specified by
$(\epsilon + 0 + 00)(1 + 10 + 100)^* 1$, and on the right
by a string specified by $(1+10+100)^+$.  Thus every string in $\Delta^*$
has one of the following forms:
\begin{itemize}
\item[(a)] $y$
\item[(b)] $yw$
\item[(c)] $(zx)^* z$
\item[(d)] $yx(zx)^* z$
\item[(e)] $(zx)^* zw$
\item[(f)] $yx (zx)^* z w$
\end{itemize}
where $y = \lbrace \epsilon, 0, 00 \rbrace$,
$x = \lbrace 1, 10, 100 \rbrace^* 1$, 
$z = \lbrace 000 \rbrace \lbrace 0 \rbrace^*$,
and $w = \lbrace 1, 10, 100 \rbrace^+$.
For forms (a)-(f), we argue that each string $s$ specified is accepted by
$M'$.  We do this only for part (f), as the others are similar.

Let $s \in \Delta^*$.  We show how to construct an accepting path
for $s$ in $M'$, where $s$ is of the form
$yx (zx)^* z w$.  Write $s = y' x_0 z_0 x_1 \cdots z_{n-1} x_{n-1} z_n w'$,
where $y'$ is a string of $y$, each $x_i$ is a string of $x$, each
$z_i$ is a string of $z$, and $w'$ is a string of $w$.

First, consider an accepting path for $2h^{-1}(x_0)$ in $M$.
This path corresponds to a path in $M'$ starting at $q_0$ and
visiting a sequence of old states in turn.
In particular, the path for the prefix $2$ 
corresponds in $M'$ to a sequence of transitions on (successively) $1,0,0$,
leading to an old state.  Call the sequence of states encountered
$q_1$, $q_2$, $q_3$.  Since every state of $M'$ is initial, we can
choose to start at our accepting path at 
\begin{itemize}
\item $q_1$ (if the string $s$ we are trying to accept starts with $001$);
\item $q_2$ (if the string $s$ we are trying to accept starts with $01$);
\item $q_3$ (if the string $s$ we are trying to accept starts with $1$).
\end{itemize}
Thus there is a path in $M'$ starting at either $q_1$, $q_2$, or $q_3$,
processing $y' x_0 $, and ending in an old state.  At this point we can read
$z_0$, which leads back to $q_0$.   It now remains
to construct a path for $x_1 z_1 \cdots x_{n-1} z_n w'$.
Again, there is path from $q_0$ in $M$ on $h^{-1} (x_1)$, and this
corresponds to a path in $M'$ leading to an old state.  We can now
process the symbols of $z_1$, leading back to $q_0$.  This process
continues until after reading $z_n$ we have returned once more to $q_0$.
At this point we can process the symbols of $w'$, and we are in an
accepting state.  Thus $M'$ accepts $s$.







For the other direction,
assume $M'$ accepts $\Delta^*$.  We must show $M$ accepts $\Sigma^*$.
Clearly $M$ accepts $\epsilon$, since $M$ has an initial state and all
states are final.  Now let $s \in \Sigma^+$, and consider the string
$1000h(s)1$ in $\Delta^*$.  This string is accepted, and so there is an
accepting path starting in some state (not necessarily $q_0$) for it in
$M'$.  By our construction,
after reading $1000$, we are either in $q_0$ or in some new state.
If we are in a new state, however, there is no possible transition
on $1$, so we must be in $q_0$ after reading $000$.  Now an acceptance
path for $h(s)1$ from $q_0$ corresponds to an acceptance path for
$s0$, and hence $s$, in $M$.  (We require the final $1$ because otherwise
if $s$ ends in $0$, we could be in a new state of $M'$ which would not
map back to a path in $M$.)
\end{proof}

This completes the proof of Theorem~\ref{asif}.
\end{proof}

\begin{corollary}
Minimizing an NFA with all states both initial and final is 
PSPACE-hard.
\end{corollary}

%
%
%

\section{Characterization of the languages accepted by special NFAs}
\label{charac}

In this section we observe that the languages accepted by the kinds of
NFAs we have been discussing have a simple characterization.

We define $\pref(L)$ to be the language of all prefixes of strings of
$L$, $\suff(L)$ to be the language of all suffixes of strings of $L$, and
$\subword(L)$ to be the language of all factors (aka ``subwords'') of
strings of $L$.  A language $L$ is {\it prefix-closed} if $L= \pref(L)$,
{\it suffix-closed} if $L = \suff(L)$,
and {\it factorial} if $L = \subword(L)$.

The results summarized in the following theorem are easy to prove.
Part~(b) was noted by Gill and Kou \cite{Gill&Kou:1974}.

\begin{theorem}
 	\begin{itemize}
 	\item[(a)] A nonempty regular language is prefix-closed
 	if and only if it is accepted by some NFA with all states final;

 	\item[(b)] A nonempty regular language is suffix-closed
 	if and only if it is accepted by some generalized
 	NFA with all states initial
 	and one final state.

 	\item[(c)] A nonempty regular language is factorial if and only if it is
 	accepted by some generalized NFA with all states both initial
 	and final.
 	\end{itemize}
\end{theorem}



It is natural to consider the complexity of testing whether a given
regular language is prefix-closed, suffix-closed, or factorial.  We
will see below that the answer depends on whether the input is given
as an NFA or a DFA.

\begin{theorem}
The following problems are PSPACE-complete: given an NFA $M$, decide
if $L(M)$ is not prefix-closed (resp. suffix-closed, factorial).
\end{theorem}

\begin{proof}
To show that determining if $L(M)$ is not prefix-closed is in PSPACE, we first
give a non-deterministic algorithm.  The desired result will then follow by
Savitch's Theorem.  Let $n$ be the number of states of $M$.
If $L(M)$ is not prefix-closed, there exists a string $w \in L(M)$
such that some prefix $w'$ of $w$ is not in $L(M)$.
We guess such a $w$ of length $< 2^{n+1}$ one input symbol at a time
and verify that $w$ is accepted by $M$ but some prefix $w'$ is not.
The space required is that for the current set of states of $M$
and for an $n+1$ bit counter, which is clearly polynomial.
It remains to show that if such a $w$ exists, we may choose $w$ to
have length $< 2^{n+1}$.  Suppose the shortest such $w$ has length
$\geq 2^{n+1}$.  Let $w'$ be the prefix of $w$ not accepted by $M$.
During the computation of $M$ on the first $2^n$ symbols of $w$, $M$
must repeat a set of states, and similarly for its computation
on the second $2^n$ symbols of $w$.  If $w'$ has length $> 2^n$,
then omitting the portion of the computation between the
repeated set of states in the first half of $w$ yields a new, shorter
string accepted by $M$ with a prefix not accepted by $M$,
contradicting the minimality of $w$.  If
$w'$ has length $\leq 2^n$, then omitting the portion of the computation
between the repeated set of states in the second half of $w$ gives
the same result.  We conclude that a shortest such $w$ has length $< 2^{n+1}$.

A similar argument shows that determining if $L(M)$ is not suffix-closed
is also in PSPACE.  Noting that $\subword(L) = \suff(\pref(L))$, one
concludes that determining if $L(M)$ is factorial is also in PSPACE.

To show PSPACE-hardness we use the reduction from the acceptance
problem for polynomial-space bounded Turing machines to
{\tt NFA-NONUNIVERSALITY} given by Aho, Hopcroft, and Ullman
\cite[Section~10.6]{Aho&Hopcroft&Ullman:1974}.
Given a deterministic Turing machine $T$ and an input $w$,
Aho, Hopcroft, and Ullman \cite[Section~10.6]{Aho&Hopcroft&Ullman:1974}
showed how to construct a regular expression $E$ specifying 
all strings that do not represent
an accepting computation of $T$ on $w$.  From $E$ we can construct
an NFA $M$ for $L(E)$ in polynomial space using the standard constructions.
Thus if $T$ does not accept
$w$, the NFA $M$ accepts all strings over its input alphabet $\Sigma$.  If
$T$ does accept $w$, then $M$ accepts all strings except the one
string $x$ that represents the accepting computation of $T$
on $w$.  But now if $L(M) = \Sigma^*$, then $L(M)$ is clearly
prefix-closed, suffix-closed, and factorial.  If
$L(M) = \Sigma^* \setminus \{x\}$, then $L(M)$ is not prefix-closed,
suffix-closed, or factorial.  Thus $L(M) = \Sigma^*$ iff
$L(M)$ is prefix-closed (resp. suffix-closed, factorial).
Since the problem of deciding if $L(M) \neq \Sigma^*$ is PSPACE-complete,
we conclude that deciding if $L(M)$ is not prefix-closed (resp. suffix-closed,
factorial) is PSPACE-complete.
\end{proof}

\begin{theorem}
The following problems can be solved in polynomial time: given a DFA $M$,
decide if $L(M)$ is not prefix-closed (resp. suffix-closed, factorial).
\end{theorem}

\begin{proof}
Given a DFA $M$ we may easily construct a DFA $M'$ accepting $\pref(L)$
by making final every state in $M$ that can reach a final state.
To test if $L(M)$ is not prefix-closed is to test the non-emptiness
of $L(M') \setminus L(M)$, which is easily done in polynomial time
by the cross-product construction and the standard
algorithm for testing the emptiness of a language accepted by a DFA.

To determine if $L(M)$ is not suffix-closed, first let
$M = (Q,\Sigma,\delta,0,F)$, where $Q = \{0,\ldots,n-1\}$.
We construct at most $n$ new DFAs $M_i$, $0 \leq i \leq n-1$, where
$i$ is a state of $M$ reachable from $0$ and $M_i$ is identical to $M$
except that $i$ is the start state of $M_i$.  We now test if any of the
$M_i$ accept a string not accepted by $M$.  As before, this can be done
in polynomial time for each $M_i$, and since we have at most $n$ machines
$M_i$, the overall runtime is polynomial.

To determine if $L(M)$ is not factorial, we construct the $M_i$ as above,
but now for each $M_i$ we make final every state of $M_i$ that can reach
a final state.  Again, we now test if any of the $M_i$ accept a string
not accepted by $M$.
\end{proof}

For more exact analysis of the running time, see \cite{Brzozowski:2008}.

\section{State complexity results}
\label{statecomp}

       We now turn to state complexity results.  It is well known that, for all $n \geq 1$,
there exists an NFA with $n$ states such that minimal equivalent DFA
has $2^n$ states.  In this section we show that the maximum blow-up can still be achieved
for alphabets of size $\geq 2$,
if we demand that all states be final, initial,
or both initial and final.  We note that
in computing the state complexity, we demand that our DFAs be complete,
that is, that there is a well-defined transition
from every state and every input symbol.

      The situation is somewhat different for the unary case, with alphabet
$\Sigma=\lbrace a \rbrace$.   In the case
of an NFA with all final states, the maximum blow-up in going from an
NFA to a DFA is $n \rightarrow n+1$ states.  To see this, note that if
a unary $n$-state NFA with all final states has a directed cycle,
then it accepts
$a^*$, which can be done with a $1$-state DFA.  Otherwise there 
exists a $k \leq n$ such that $a^k$ is the shortest string not accepted.
This can be accepted with a $k+1$-state DFA (by adding the missing
dead state).  In the case $k = n$, this results in a $n \rightarrow n+1$
blowup.  The same results occur for NFAs with all states initial and
one final, or with all states both initial and final.

      Now we turn to the case of larger alphabets.

\begin{theorem}
\label{allfinal}
For $n = 1$ and every $n \geq 3$
there exists an NFA $M$ over a binary alphabet with $n$ states, all
of which are final, such that the minimal DFA accepting $L(M)$
has $2^n$ states.  No such binary NFA exists for $n = 2$, although over
a ternary alphabet one exists.
\end{theorem}

\begin{proof}
For $n = 1$ we take the automaton with a single state which is both initial
and final, with a self-loop on only one of the two letters.

For $n=2$ we can enumerate all possible binary NFAs with all states
final and check that none of them
have a minimal DFA with $4$ states.

It is easy to verify
that the ternary NFA in Figure~\ref{nfa6}
has deterministic state complexity $4$.

\begin{figure}[H]
\centering
\includegraphics[scale=1.00]{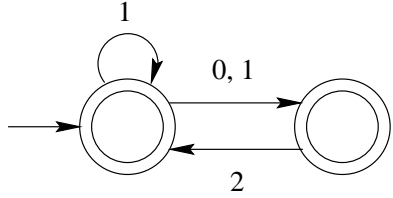}
\caption{A two-state NFA with both states final where the minimal equivalent DFA has 4 states.}
\label{nfa6}
\end{figure}

Now assume $n \geq 3$.
We define an NFA $M = (Q,\Sigma,\delta,0,F)$ (Figure~\ref{dfa1}), where
$Q = \lbrace 0,\ldots,n-1 \rbrace$, $\Sigma = \lbrace 0,1 \rbrace$,
$F = Q$, and for any $i$, $0 \leq i \leq n-1$,
\[
\delta(i,a)=
\begin{cases}
\{i+1\}, & \text{if $a=0$ and $0 \leq i \leq n-3$;}\\
\{n-1\}, & \text{if $a=0$ and $i=n-1$;}\\
\{0,i+1\}, & \text{if $a=1$ and $0 \leq i \leq n-2$.}
\end{cases}
\]

\begin{figure}[ht]
\centering
\includegraphics{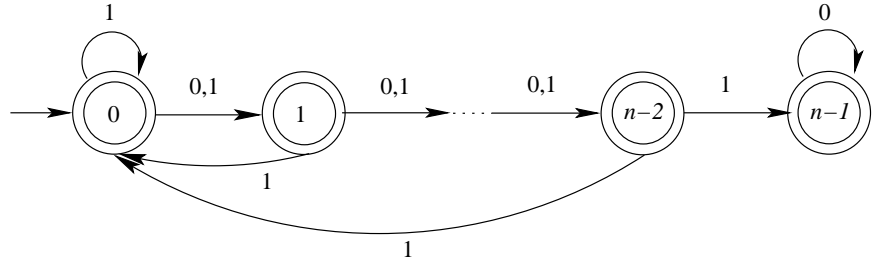}
\caption{The NFA $M$ of Theorem~\ref{allfinal}.}
\label{dfa1}
\end{figure}

Let $M' = (2^Q,\Sigma,\delta',\{0\},F')$ be the DFA obtained by
applying the subset construction to $M$.
To show that $M'$ is minimal we will show (a) that all states of $M'$
are reachable, and (b) that the states of $M'$ are pairwise
inequivalent with respect to the Myhill--Nerode equivalence
relation.

To prove part~(a) let $S \subseteq Q$ be a state of $M'$, where
$S = \{ s_1, s_2, \ldots , s_k \}$ for some $k$ and
$s_1 < s_2 < \cdots < s_k$.  There are two cases to consider.

Case 1: $n-1 \notin S$.  Then
\[
\delta'(\{0\},0^{s_k - s_{k-1} - 1}1 0^{s_{k-1} - s_{k-2} - 1}1 \cdots
0^{s_2 - s_1 - 1}1 0^{s_1}) = S.
\]

To see this, let $w_k = \epsilon$ and for $1 \leq i \leq k-1$, let
\[
w_i = 0^{s_k - s_{k-1} - 1}1 0^{s_{k-1} - s_{k-2} - 1}1 \cdots
0^{s_{i+1} - s_{i} - 1}1.
\]
For $1 \leq i \leq k$, let $S_i = \delta'(\{0\},w_i)$.
Then $S_{i-1} = S_i + s_i - s_{i-1} \cup \{0\}$.
We see that $S_i = \{s_i < s_{i+1} < \cdots < s_k\} - s_i$.
Here, for $m \in Q$, the notation $S + m$ refers to the set
$\lbrace x+m : x \in S \rbrace$.
Thus
\[
\delta'(\{0\},w_1 0^{s_1}) = \delta'(S_1,0^{s_1}) = S_1 + s_1 = S,
\]
as required.

Case 2: $n-1 \in S$.  By the argument of Case~1, $S \setminus \{ n-1 \}$
is reachable.  But then
\[
\delta'(S \setminus \{n-1\},0^{n - 2 - s_{k-1}}1
0^{s_{k-1} - s_{k-2} - 1}1 \cdots 0^{s_2 - s_1 - 1}1 0^{s_1}) = S.
\]

To see this, for $1 \leq i \leq k-1$, let
\[
w_i = 0^{n - 2 - s_{k-1}}1 0^{s_{k-1} - s_{k-2} - 1}1 \cdots
0^{s_{i+1} - s_{i} - 1}1.
\]
For $1 \leq i \leq k-1$, let $S_i = \delta'(S \setminus \{n-1\},w_i)$.
Then
\[
S_{i-1} = ((S_i \setminus \{n-1\}) + s_i - s_{i-1}) \bmod{(n-1)}) \cup \{n-1\}.
\]
We see that
\begin{eqnarray*}
S_i &=& ((S \setminus \{n-1\}) + n - 1 - s_i) \bmod{(n-1)}) \cup \{n-1\}\\
&=& ((S \setminus \{n-1\}) - s_i) \bmod{(n-1)}) \cup \{n-1\}.
\end{eqnarray*}
Thus
\begin{eqnarray*}
\delta'(S \setminus \{n-1\},w_1 0^{s_1}) &=& \delta'(S_1,0^{s_1})\\
&=& ((S \setminus \{n-1\}) - s_i + s_i) \bmod{(n-1)}) \cup \{n-1\} \\
&=& S \setminus \{n-1\} \cup \{n-1\}\\
&=& S,
\end{eqnarray*}
as required.

To prove part~(b) let $S$ and $T$ be distinct states of $M'$.
We have 2 cases.

Case 1: $n-1$ is in exactly one of $S$ or $T$.  Without loss of generality,
suppose $n-1 \notin S$ and $n-1 \in T$.  Then $\delta'(S,0^{n-1}) = \emptyset$
and $\delta'(T,0^{n-1}) = \{n-1\}$, so $S$ and $T$ are inequivalent.

Case 2: either $n-1$ is in both of $S$ and $T$ or $n-1$ is in neither.
Without loss of generality, suppose there exists $i \notin S$, $i \in T$.
Then $\delta'(S,0^{n-2-i}1) = S'$ and $\delta'(T,0^{n-2-i}1) = T'$,
where $n-1 \notin S'$ and $n-1 \in T'$.  We now apply the argument of Case 1.
\end{proof}

     We now turn to the case where all states are both initial and final.

\begin{theorem}
\label{allinitial}
For every $n \geq 1$ there exists an NFA $M$ over a binary
alphabet with $n$ states, each
of which is both initial and final, such that the minimal DFA accepting $L(M)$
has $2^n$ states.
\end{theorem}

\begin{proof}
For $n = 1$ we take the automaton with a single state which is both initial
and final, with a self-loop on only one of the two letters.

Now assume $n \geq 2$.
We define an NFA $M = (Q,\Sigma,\delta,Q,F)$ (Figure~\ref{dfa2}), where
$Q = \lbrace 0,\ldots,n-1 \rbrace$, $\Sigma = \lbrace 0,1 \rbrace$,
$F = Q$, and for any $i$, $0 \leq i \leq n-1$,
\[
\delta(i,a)=
\begin{cases}
\{i-1\}, & \text{if $a=0$ and $1 \leq i \leq n-1$;}\\
\{n-1\}, & \text{if $a=0$ and $i=0$;}\\
\{i+1\}, & \text{if $a=1$ and $0 \leq i \leq n-2$.}
\end{cases}
\]

\begin{figure}[ht]
\centering
\includegraphics{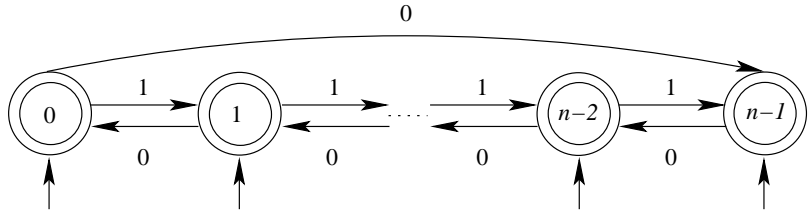}
\caption{The NFA $M$ of Theorem~\ref{allinitial}.}
\label{dfa2}
\end{figure}

Let $M' = (2^Q,\Sigma,\delta',Q,F')$ be the DFA obtained by
applying the subset construction to $M$.
To show that $M'$ is minimal we will show (a) that all states of $M'$
are reachable, and (b) that the states of $M'$ are pairwise
inequivalent with respect to the Myhill--Nerode equivalence
relation.

To prove part~(a) let $S \subseteq Q$ be a state of $M'$, where
$S = Q \setminus \{ s_1, s_2,  \ldots,  s_k \}$ for some $k$ and
$s_1 < s_2 < \cdots < s_k$.  Then
\[
\delta'(Q,0^{s_1 + 1}1 0^{s_2 - s_1 + 1}1 \cdots 0^{s_k - s_{k-1} + 1}1
0^{n - s_k}) = S.
\]

To see this, for $1 \leq i \leq k$, let
\[
w_i = 0^{s_1 + 1}1 0^{s_2 - s_1 + 1}1 \cdots 0^{s_i - s_{i-1} + 1}1,
\]
and let $S_i = \delta'(Q,w_i)$.
Then one easily verifies that
\[
S_i = ((Q \setminus \{s_1,  s_2,  \ldots,  s_i\}) - s_i) \bmod{n},
\]
so $\delta'(Q,w_k 0^{n-s_k}) = \delta'(S_k,0^{n-s_k}) =
(S - s_k - n + s_k) \bmod{n} = S$.

To prove part~(b) let $S$ and $T$ be distinct states of $M'$.
Without loss of generality, suppose there exists $i \notin S$, $i \in T$.
Then $\delta'(S,0^i 1^{n-1}) = \emptyset$ and
$\delta'(T,0^i 1^{n-1}) = \{n-1\}$, so $S$ and $T$ are inequivalent.
\end{proof}

Finally, we consider the case where all states are initial, and only
one state is final.  An example of maximal blowup from $n$ states
to $2^n$ deterministic states was first given by Gill and Kou
\cite{Gill&Kou:1974}, but their construction was not over a fixed
alphabet.  Later, Veloso and Gill \cite{Veloso&Gill:1979}
gave an example over a binary alphabet.  Here we give another example.
The following NFA, which is a trivial variation
on that in Figure~\ref{dfa2}, demonstrates the maximum blow-up from $n$
states to $2^n$ deterministic states for all $n \geq 1$.
We omit the proof, which is
a trivial variation of the proof of Theorem~\ref{allinitial}.

\begin{figure}[ht]
\centering
\includegraphics{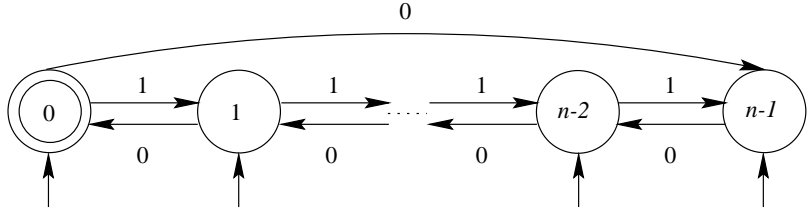}
\caption{The NFA demonstrating maximum blow-up for all states initial, one state
final.}
\label{dfa10}
\end{figure}

\section{State complexity of $\pref(L)$, $\suff(L)$, $\subword(L)$}
\label{statecomp2}

      In this section we consider the the state complexity of the operations
$\pref(L)$, $\suff(L)$, and $\subword(L)$.

       If the state complexity of $L$ is $n$, the
state complexity of $\pref(L)$ is also at most $n$, as can be seen from
the standard construction for $\pref(L)$ where we change every state from
which a final state can be reached to final.

       The state complexity of $\suff(L)$ is more interesting.

\begin{theorem}
\label{suff}
Let $M$ be a DFA with $n$ states.  Then $\suff(L(M))$ can be accepted
by a DFA with at most $2^n - 1$ states, and this bound is tight.
\end{theorem}

\begin{proof}
Let $M = (Q,\Sigma,\delta,0,F)$, where $Q = \{0,\ldots,n-1\}$.
Then $\suff(L(M))$ is accepted by the generalized NFA
$N = (Q,\Sigma,\delta,P,F)$, where $P \subseteq Q$ is the set
of states reachable from the start state.  But it is clear that the empty
set is not reachable from any nonempty set of states of $N$, so the minimal
equivalent DFA has at most $2^n - 1$ states.

To show the bound is tight, consider the DFA $M = (Q,\Sigma,\delta,0,F)$
on states $Q = \lbrace 0, 1, \ldots, n-1 \rbrace$ (Figure~\ref{dfa3})
defined by
\begin{eqnarray*}
\delta(q, 0) &=& q , \ \text{for $0 \leq q < n-1$;} \\
\delta(n-1, 0) &=& 0; \\
\delta(q,1) &=& (q+1) \bmod n, \ \text{for $0 \leq q < n$;}
\end{eqnarray*}
and with $F = \lbrace 0 \rbrace $.

\begin{figure}[ht]
\centering
\includegraphics{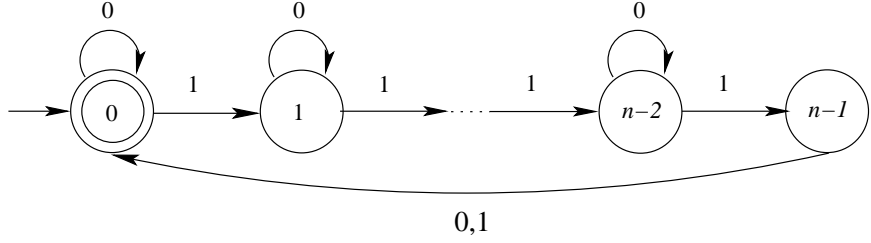}
\caption{The DFA $M$ of Theorem~\ref{suff}.}
\label{dfa3}
\end{figure}

Now consider the generalized NFA $N = (Q,\Sigma,\delta,Q,F)$.
By the argument above, $N$ accepts $\suff(L(M))$.
Let $M' = (2^Q \setminus \emptyset,\Sigma,\delta',Q,F')$ be the DFA
obtained by applying the subset construction to $N$ and
removing the empty set.
To show that $M'$ is minimal we will show (a) that all states of $M'$
are reachable, and (b) that the states of $M'$ are pairwise
inequivalent with respect to the Myhill--Nerode equivalence
relation.

To prove part~(a) let $S \subseteq Q$ be a state of $M'$, where
$S = Q \setminus \{ s_1,  s_2, \ldots, s_k \}$ for some $k$ and
$s_1 < s_2 < \cdots < s_k$.
Let $T \subseteq Q$, $T \neq \emptyset$.  If both $t$ and $t+1$
are in $T$, $t < n-1$, then one easily verifies that
\[
\delta'(T,1^{n-1-t}01^{t+1}) = T \setminus \{t\}.
\]
For $0 \leq i < n-1$, define $w_i = 1^{n-1-i}01^{i+1}$.  We have two cases.

Case 1: $s_k \neq n-1$.  We see that
\[
\delta'(Q,w_{s_1}w_{s_2}\cdots w_{s_k}) =
Q \setminus \{ s_1,  s_2,  \ldots, s_k \} = S,
\]
as required.

Case 2: $s_k = n-1$.  Let $T = Q \setminus
\{ s_1,  s_2,  \ldots,  s_{k-1} \}$, where
$s_1 < s_2 < \cdots < s_{k-1}$.  By the argument of Case 1
\[
\delta'(Q,w_{s_1}w_{s_2}\cdots w_{s_{k-1}}) = T.
\]
Since $S \neq \emptyset$, there exists a smallest $t \in T$, $t \neq n-1$.
If $t = 0$, then $\delta'(T,0) = T \setminus \{n-1\} = S$.  Otherwise,
\[
\delta'(T,0(1^{n-1}0)^{t-1}1^{t-1}) = T \setminus \{n-1\} \cup \{t-1\} = T'.
\]
But now
\[
\delta'(T',w_{t-1}) = T \setminus \{n-1\} = S,
\]
as required.

To prove part~(b) let $S$ and $T$ be distinct states of $M'$.
Without loss of generality, suppose there exists $i \notin S$, $i \in T$.
The set of final states $F'$ consists of all subsets of $Q$ containing
$0$.  But $0 \notin \delta'(S,1^{n-i})$ and $0 \in \delta'(T,1^{n-i})$,
so $S$ and $T$ are inequivalent.
\end{proof}

     We now turn to the state complexity of $\subword(L)$:

\begin{theorem}
\label{subword}
Let $M$ be a DFA with $n$ states.  Then $\subword(L(M))$ can be 
accepted by a DFA with at most $2^{n-1}$ states, and this
bound is tight.
\end{theorem}

\begin{proof}
Let $M = (Q,\Sigma,\delta,0,F)$, where $Q = \{0,\ldots,n-1\}$.
Let us assume that $M$ contains no unreachable states.
Suppose that every state of $M$ can reach a final state.
Then $\subword(L(M)) = \Sigma^*$ and is accepted by a one state DFA.
Let us suppose then that there exists $q \in Q$ such that $q$
cannot reach a final state.  Then we may remove the state $q$
and any associated transitions to obtain a equivalent NFA with
$n-1$ states.  Then $\subword(L(M))$ is accepted by the generalized NFA
$N = (Q \setminus \{q\},\Sigma,\delta,Q \setminus \{q\},P)$, where
$P \subseteq Q$ is the set of states that can reach a final state.
The minimal DFA equivalent to $N$ thus has at most $2^{n-1}$ states.

To show the bound is tight, consider the DFA $M$ on states
$Q = \lbrace 0, 1, \ldots, n-1 \rbrace$ (Figure~\ref{dfa4}) defined by
\begin{eqnarray*}
\delta(q,0) &=& q , \ \text{for $0 \leq q < n-2$;} \\
\delta(q,0) &=& n-1 , \ \text{for $q = n-2, n-1$;} \\
\delta(q,1) &=& (q+1) \bmod n, \ \text{for $0 \leq q < n-2$;} \\
\delta(n-2,1) &=& 0; \\
\delta(n-1,1) &=& n-1 ;
\end{eqnarray*}
and with $F = \lbrace 0 \rbrace $.

\begin{figure}[ht]
\centering
\includegraphics{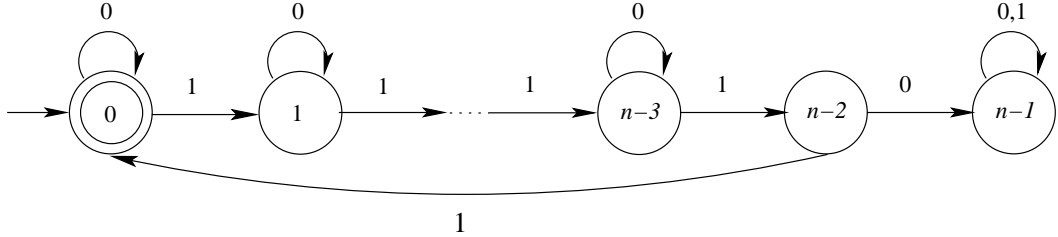}
\caption{The DFA $M$ of Theorem~\ref{subword}.}
\label{dfa4}
\end{figure}

Note that state $n-1$ cannot reach a final state.  Let $\tilde{M}$ be
the NFA obtained by removing state $n-1$ from $M$, along with all
associated transitions.  Let $N$ be the generalized NFA obtained from
$\tilde{M}$ by making all states both initial and final.  Then $N$
accepts $\subword(L(M))$.  Let $Q' = \{0,\ldots,n-2\}$.  Let
$M' = (2^{Q'},\Sigma,\delta',Q',F')$ be the DFA obtained by
applying the subset construction to $N$.  To show that $M'$ is minimal
we will show (a) that all states of $M'$ are reachable, and (b) that
the states of $M'$ are pairwise inequivalent with respect to the
Myhill--Nerode equivalence relation.

To prove part~(a) let $S \subseteq Q'$ be a state of $M'$, where
$S = Q' \setminus \{ s_1,  s_2,  \ldots,  s_k \}$ for some $k$, and
$s_1 < s_2 < \cdots < s_k$.
One easily verifies that for any $T \subseteq Q'$ and $t \in Q'$,
\[
\delta'(T,1^{n-2-t}01^{t+1}) = T \setminus \{t\},
\]
from which it is clear that $S$ is reachable.

To prove part~(b) let $S$ and $T$ be distinct states of $M'$.
Without loss of generality, suppose there exists $i \notin S$, $i \in T$.
Then by the argument of part~(a), there exists a string $w$ such
that $\delta'(S,w) = \emptyset$ and $\delta'(T,w) = \{i\}$,
so $S$ and $T$ are inequivalent.
\end{proof}


\section{Nondeterministic state complexity of complement}
\label{complement}

      We now consider the following question.  Let $M$ be an NFA
with all states final, accepting a language $L$.
What is the maximum size of a minimal NFA
accepting $\overline{L}$?

      The case where we remove the restriction that all states be
final was previously studied by Sakoda and Sipser
\cite{Sakoda&Sipser:1978}, Birget \cite{Birget:1993},
Ellul et al.\ \cite{Ellul&Krawetz&Shallit&Wang:2004}, and
Jir\'askov\'a \cite{Jiraskova:2005}.  Jir\'askov\'a
constructed an $n$ state NFA $N$ over the alphabet $\{0,1\}$
such that any NFA accepting $\overline{L(N)}$ requires
at least $2^n$ states.

Jir\'askov\'a's NFA is defined as follows:  let
$N = (Q,\Sigma,\delta,0,F)$, where $Q = \{0,\ldots,n-1\}$,
$\Sigma = \{0,1\}$, $F = \{n-1\}$, and for any $i$, $0 \leq i \leq n-1$,
\[
\delta(i,a)=
\begin{cases}
\{0,i+1\}, & \text{if $a=0$ and $i < n-1$;}\\
\{1,2,\ldots,n-1\}, & \text{if $a=0$ and $i = n-1$;}\\
\{i+1\}, & \text{if $a=1$ and $i < n-1$.}
\end{cases}
\]

By modifying this construction we prove

\begin{theorem}
For $n \geq 1$, there exists an NFA $M$ of $n+1$ states over a three-letter
alphabet with all states final such that any NFA accepting $\overline{L(M)}$
requires at least $2^n$ states.
\end{theorem}

\begin{proof}
Let $N$ be the NFA described above.  Let $Q' = \{0,\ldots,n\}$
and let $M = (Q',\{0,1,2\},\delta',0,Q')$, where for any $i$,
$0 \leq i \leq n$,
\[
\delta'(i,a)=
\begin{cases}
\delta(i,a), & \text{if $a \neq 2$ and $i < n$;}\\
\{n\}, & \text{if $a=2$ and $i = n-1$.}
\end{cases}
\]
Then by modifying the fooling set argument of Jir\'askov\'a
\cite[Theorem~5]{Jiraskova:2005} one obtains a fooling set
of size $2^n$ for $\overline{L(M)}$, giving the desired result.  (One
obtains the fooling set for $\overline{L(M)}$ by appending a $2$ to the
second word in each pair of the fooling set for $\overline{L(N)}$.)
\end{proof}

\section{Shortest word not accepted}
\label{shortest}

     Finally, we consider one more problem.  Given an $n$-state NFA $M$ with
all states final, such that $L(M) \not= \Sigma^*$, how long can the
shortest unaccepted string be?     At first glance it might appear that
such a string has to be of length $\leq n$, but this is not the case.

\begin{theorem}
There exists an $n$-state NFA $M$ with all states final, such that
the smallest string not accepted by $M$ has length $2^{cn}$ for some
constant $0 < c \leq 1$.
\end{theorem}

\begin{proof}
In \cite{Ellul&Krawetz&Shallit&Wang:2004} the authors show that there exist
$n$-state NFAs $M$ over a $2$-letter alphabet $\Sigma$
such that the shortest string
not accepted is of length $2^{cn}$ for some constant $0 < c \leq 1$.
We take such an NFA $M$, and add a new symbol, say $\#$, with transitions
on $\#$ from every final state of $M$ back to $M$'s initial state.   Now make
all states final.  Call the resulting NFA $M'$.
Since $M$ accepts $\epsilon$, its
initial state is also final, and hence $M'$ has a transition from its
initial state to itself on $\#$.

      We claim that $L(M') \not= (\Sigma \ \cup \lbrace \# \rbrace)^*$,
but the shortest string
not accepted by $M'$ is at least as long as that for $M$.  Let $w$ be 
the shortest string not
accepted by $M$, of length $N$.
Then either there is no path in $M$ labeled $w$, or
every path labeled $w$ in $M$, arrives at a non-accepting state in $M$.
In either case $w\#$ fails to be accepted by $M'$.  On the other hand,
$M'$ accepts all strings shorter than $w$,
since any shorter string $w'$ is of the form
$w_1 \# w_2 \# \cdots \# w_r$ for some strings $w_1, w_2, \ldots, w_r \in
\Sigma^*$, where each $w_i$ has length $< N$.  Starting in the initial
state of $M'$, we read $w_1$, which is accepted by $M$ since it is of
length $< |w|$.  If $w' = w_1$, then $w_1$ is accepted by $M'$.  Otherwise
we follow the transition on $\#$ back to the initial state of $M'$ and
continue with $w_2$, etc.
\end{proof}

      We can obtain a similar result for NFAs where
all states are both initial and final.
In this case, we again add a new symbol $\#$, with transitions
on $\#$ from every final state of $M$ back to $M$'s initial state, and then
make all states both initial and final.  Now we argue as
above, except we consider the string $\# w \#$ instead.

\section{Acknowledgments}

     We are very grateful to
Andrew Malton for having suggested the problem.  We thank
the referees for reading the paper carefully and correcting several errors.

\end{document}